\documentclass[12pt]{article}
\usepackage{amsmath,amsfonts}
\usepackage{graphicx}
\usepackage{psfrag}

\numberwithin{equation}{section}

\begin{document}
\newcommand{\todo}[1]{{\em \small {#1}}\marginpar{$\Longleftarrow$}}   
\newcommand{\labell}[1]{\label{#1}\qquad_{#1}} 
\newcommand{\ud}{\mathrm{d}}

\rightline{DCPT-05/17}   
\rightline{hep-th/0504181}   
\vskip 1cm 

\begin{center} 
{\Large \bf Non-supersymmetric smooth geometries and D1-D5-P bound
states}
\end{center} 
\vskip 1cm   
  
\renewcommand{\thefootnote}{\fnsymbol{footnote}}   
\centerline{\bf Vishnu Jejjala\footnote{vishnu.jejjala@durham.ac.uk},
  Owen Madden\footnote{o.f.madden@durham.ac.uk}, Simon 
F. Ross\footnote{S.F.Ross@durham.ac.uk} and Georgina
Titchener\footnote{G.L.Titchener@durham.ac.uk}}     
\vskip .5cm   
\centerline{ \it Centre for Particle Theory, Department of  
Mathematical Sciences}   
\centerline{\it University of Durham, South Road, Durham DH1 3LE, U.K.}   
  
\setcounter{footnote}{0}   
\renewcommand{\thefootnote}{\arabic{footnote}}

\vskip 1cm
\begin{abstract}
We construct smooth non-supersymmetric soliton solutions with
D1-brane, D5-brane and momentum charges in type IIB supergravity
compactified on $T^4 \times S^1$, with the charges along the compact
directions. This generalises previous studies of smooth supersymmetric
solutions. The solutions are obtained by considering a known family of $U(1)
\times U(1)$ invariant metrics, and studying the conditions imposed by
requiring smoothness. We discuss the relation of our solutions to
states in the CFT describing the D1-D5 system, and describe various
interesting features of the geometry.
\end{abstract}
\newpage

\section{Introduction and Summary}

String theory has made tremendous advances in understanding the
microscopic origins of black hole entropy \cite{sen, sv}.  In the
original calculations, two different dual descriptions of a
supersymmetric object were considered: a weakly-coupled description in
terms of perturbative strings and D-branes, and a strongly-coupled
description as a classical black hole solution. The picture of this
black hole, as a background for the perturbative string, is
essentially the same as in semiclassical general relativity.  We have
a singularity in spacetime that is shielded (censored) by a horizon.
The horizon area determines the Bekenstein-Hawking entropy $S_{BH} =
\frac{A}{4G_N}.$ This entropy was successfully reproduced by counting
the degenerate supersymmetric vacua in the dual perturbative D-brane
description. This picture did not provide any understanding of where
the microstates were in the strong-coupling black hole picture: smooth
black hole solutions `have no hair', so the geometry is entirely
determined by the charges~\cite{nohair}. There was, however, a
suggestion that pure states would be dual to geometries which were not
smooth at the event horizon~\cite{myers}.

The anti-de Sitter/conformal field theory (AdS/CFT)
correspondence~\cite{m, gkp, w} provided a deeper understanding of the
counting of black hole entropy in string theory. The black holes in
AdS are identified with the thermal ensemble in the dual CFT. The CFT
was conjectured to provide a fundamental, non-perturbative description
of string theory with asymptotically AdS boundary conditions, so the
microstates were fundamentally thought of as states in the CFT, and it
did not appear that they could be thought of as living somewhere in
the black hole geometry. The evolution of the states in the CFT is
unitary. Certain states can be identified with classical geometries,
but as has been emphasised in e.g.~\cite{w, eternal}, the CFT provides
a fully quantised description, and reproducing the behaviour of the
CFT from a spacetime point of view will in general involve a sum over
bulk geometries.

In a series of papers, Mathur and his collaborators have challenged
the conventional picture of a black hole in string theory (see
\cite{mathurrev} for a review).  They argue that the black hole
geometry is merely a coarse grained description of the spacetime, and
that each of the $e^{S_{BH}}$ microstates can be identified with a
perfectly regular geometry with neither horizon nor singularity
\cite{lm2,stat}. The black hole entropy is a result of averaging over
these different geometries, which produces an `effective horizon',
which describes the scale at which the $e^{S_{BH}}$ individual
geometries start to differ from each other. They further argue that if
a system in an initial pure state undergoes gravitational collapse, it
will produce one of these smooth geometries, and the real spacetime
does not have a global event horizon, thus avoiding the information
loss paradox associated with outgoing Hawking radiation \cite{hawk}.
Thus, the idea is that stringy effects modify the geometry of
spacetime at the event horizon, rather than, as would be expected from
the classical point of view, at Planck or string distances from the
singularity. This is a radical modification of the expected
geometry. There are similarities with the correspondence principle
ideas~\cite{cp}, but unlike that picture, there is no obvious sense in
which the spacetime as seen by an infalling observer will be
different. It is difficult to see how the singularity behind the black
hole's event horizon can arise from a coarse graining over
non-singular geometries.\footnote{Although it may be that most
measurements in the dual CFT find it difficult to distinguish between
regular geometries and the conventional semi-classical picture of a
black hole~\cite{bjs}.}

The evidence for this proposal comes from the construction of smooth
asymptotically flat geometries in the D1-D5 system that can be
identified with individual microstates in the CFT on the worldvolume
of the branes. The theory considered is type IIB supergravity
compactified on $S^1 \times T^4 $ with $n_5$ D5-branes wrapping
$S^1\times T^4$, and $n_1$ D-strings wrapping the $S^1$. The
near-horizon geometry is $AdS_3\times S^3\times T^4$, and has a dual
$1+1$ dimensional CFT description with $c = 6 n_1 n_5$. The first such
geometries were constructed in~\cite{bbkr,maldmaoz}, and correspond to
the Ramond-Ramond (RR) ground state obtained by spectral flow from the
Neveu-Schwarz--Neveu-Schwarz (NSNS) vacuum state. This was
subsequently extended~\cite{lm1,lm2,lmm} to find a family of smooth
geometries corresponding to the whole family of RR ground states in
the CFT. The D1/D5 system via string dualities is the same as a system
with $n_5$ units of fundamental string winding and $n_1$ units of
fundamental string momentum on a circle.  The D1/D5 bound state
corresponds to a multi-wound F-string carrying momentum, and the
geometries are characterised as functions of the displacement of the
string in its transverse directions.  As a test of whether the
two-charge system indeed describes the correct physics, the collision
time for left- and right-moving excitations on the component string
was computed in field theory and compared to the time for graviton
absorption and re-emission in supergravity; the two are found to match
\cite{htube,lm2}. The degeneracy of RR ground states in this theory
gives a microscopic entropy which scales as $\sqrt{n_1 n_5}$; this was
found to match a suitable counting in a supertube description
in~\cite{palmermarolf,bak}.
However, this entropy is not large enough to correspond to a black
hole with a macroscopic horizon. It is therefore important to extend
the identification to states that carry a third charge $n_p$, momentum
along the string. These states have a microscopic degeneracy
$\sqrt{n_1 n_5 n_p}$, and were used in~\cite{sv} in the calculation of
the black hole entropy. Recently, Giusto, Mathur, and Saxena have
identified smooth geometries corresponding to some of these
states~\cite{gms1,gms2}, although the geometries constructed so far
correspond to very special states, the spectral flows of the RR ground
states studied earlier.\footnote{Three-charge states were previously
studied in the supertube description~\cite{stube1,stube2}
in~\cite{bk1,bena}. Other supersymmetric three-charge solutions have
been found in~\cite{bw,elring,bww}, but the regular solutions have not
yet been identified or related to CFT states.} The overall evidence
for the picture of black holes advanced by these authors is, in our
judgement, interesting but not yet compelling.

We will extend these investigations to find more general solitonic
solutions in supergravity, and to identify corresponding CFT
states. We believe that whether or not the picture of black holes
advanced by Mathur and collaborators proves to be correct, these
solitonic solutions will remain of interest in their own right. It is
particularly interesting that we can find completely smooth
non-supersymmetric solitons. These are, as far as we are aware, the
first explicit examples of this type.

We find these solutions by generalising an analysis previously carried
out for special cases in~\cite{bbkr,maldmaoz,gms1,gms2}. We consider
the nonextremal rotating three-charge black holes given in \cite{cvetic},
and systematically search for values of the parameters for which the
solution is smooth and free of singularities. We find that if we allow
non-supersymmetric solutions, there are two integers $m,n$ labelling
the soliton solutions. The previously studied supersymmetric solutions
correspond to $m = n+1$. Thus, we find new non-supersymmetric
solitons. Further solutions, some of which are smooth, can be
constructed by orbifolds of this basic family. This provides another
integer degree of freedom $k$. Some of the supersymmetric orbifolds
have not been previously studied.

We identify the basic family of smooth solutions labelled by $m,n$
with states in the CFT constructed by spectral flow from the NSNS
vacuum, with $m+n$ units of spectral flow applied on the left and
$m-n$ units of spectral flow applied on the right. We find a
non-trivial agreement between the spacetime charges in these
geometries and the expectations from the CFT point of view. This
agreement extends to the geometries constructed as orbifolds of the
basic smooth solutions. We have studied the wave equation on these
geometries, and we find that as in~\cite{gms2}, there is a mismatch
between the spacetime result, $\Delta t_{sugra} = \pi R \varrho$, and
the expectation from the CFT point of view, $\Delta t_{CFT} = \pi
R$. We believe that understanding this mismatch is a particularly
interesting issue for further development. Finally, we discuss the
appearance of an ergoregion in the non-supersymmetric solutions. We
find that the ergoregion does not lead to any superradiant scattering
for free fields. 

The existence of these non-supersymmetric solitons, and the fact that
they can be identified with states in the dual CFT, might be regarded
as further evidence for the proposed description of black
holes. However, we would advocate caution. We still find it
questionable whether we can really describe a black hole in this
way. First of all, the three-charge states described so far are very
special. The orbifolds we consider provide examples where the CFT
state is not the spectral flow of a RR ground state, but the
geometries we consider all have a $U(1) \times U(1)$ invariance.  It
is unclear whether the techniques used to date can be extended to
obtain even the geometries corresponding to spectral flows of the more
general RR ground states of~\cite{lm2,lmm}, let alone to reproduce the
full $e^{\sqrt{n_1 n_5 n_p}}$ states required to explain the black
hole entropy. The much more difficult dynamical questions --- how the
appearance of a global event horizon in gravitational collapse can
always be avoided, for example --- have not yet been
tackled. Nonetheless, the study of these smooth geometries offers a
new perspective on the relation between CFT and spacetime, and it is
interesting to see that their existence does not depend on
supersymmetry.

The remainder of the paper is organised as follows. In the next
section, we recall the metric and matter fields for the general family
of solutions we consider, and discuss the near-horizon limit which
relates asymptotically flat solutions to asymptotically AdS$_3 \times
S^3$ ones. In section~\ref{solitons}, we discuss the constraints
required to obtain a smooth soliton solution. We find that there is a
basic family of smooth solutions labelled by the radius $R$ of the
$S^1$, the D1 and D5 brane charges $Q_1, Q_5$, and two integers $m,n$.
Further solutions can be constructed as $\mathbb{Z}_k$ orbifolds of
these basic ones; they will be smooth if $m$ and $n$ are both
relatively prime to $k$. We also discuss the asymptotically AdS$_3
\times S^3$ solutions obtained by considering the near-horizon
limit. The asymptotically AdS$_3 \times S^3$ solutions corresponding
to the basic family of smooth solutions are always global AdS$_3
\times S^3$ up to some coordinate transformation. In
section~\ref{verify}, we verify that the solutions are indeed smooth
and free of closed timelike curves. In section~\ref{cft}, we identify
the corresponding states in the CFT, identifying the coordinate shift
in the global AdS$_3 \times S^3$ solutions with spectral
flow. In section~\ref{props}, we discuss the massless scalar wave
equation on these solutions, and show that the non-supersymmetric
solutions always have an ergoregion. Finally, in section~\ref{disc},
we discuss some directions for future research.

\section{General nonextremal solution}

We will look for smooth solutions as special cases of the nonextremal
rotating three-charge black holes given in \cite{cveticcorr} (uplifted to
ten-dimensional supergravity following~\cite{cvetlars}). The original
two-charge supersymmetric solutions of~\cite{bbkr,maldmaoz} were found
in this way, and the same approach was applied more recently
in~\cite{gms1,gms2} to find supersymmetric three-charge solutions. In
the present work, we aim to find all the smooth solutions within this
family.

In this section, we discuss this family of solutions in general,
writing the metric in forms that will be useful for finding and
discussing the smooth solutions. We will also discuss the relation
between asymptotically flat and asymptotically AdS$_3 \times S^3$
solutions.  We write the metric as
\begin{eqnarray} \label{3charge}
\mathrm{d}s^2&=&-\frac{f}{\sqrt{\tilde{H}_{1} \tilde{H}_{5}}}(
\mathrm{d}t^2 - \mathrm{d}y^2) +\frac{M}{\sqrt{\tilde{H}_{1}
\tilde{H}_{5}}} (s_p \mathrm{d}y - c_p
\mathrm{d}t)^2 \nonumber \\ &&+\sqrt{\tilde{H}_{1} \tilde{H}_{5}}
\left(\frac{ r^2 \mathrm{d}r^2}{ (r^2+a_{1}^2)(r^2+a_2^2) - Mr^2}
+\mathrm{d}\theta^2 \right)\nonumber \\ &&+\left( \sqrt{\tilde{H}_{1}
\tilde{H}_{5}} - (a_2^2-a_1^2) \frac{( \tilde{H}_{1} + \tilde{H}_{5}
-f) \cos^2\theta}{\sqrt{\tilde{H}_{1} \tilde{H}_{5}}} \right) \cos^2
\theta \mathrm{d} \psi^2 \nonumber \\ && +\left( \sqrt{\tilde{H}_{1}
\tilde{H}_{5}} + (a_2^2-a_1^2) \frac{(\tilde{H}_{1} + \tilde{H}_{5}
-f) \sin^2\theta}{\sqrt{\tilde{H}_{1} \tilde{H}_{5}}}\right) \sin^2
\theta \mathrm{d} \phi^2 \nonumber \\ && +
\frac{M}{\sqrt{\tilde{H}_{1} \tilde{H}_{5}}}(a_1 \cos^2 \theta
\mathrm{d} \psi + a_2 \sin^2 \theta \mathrm{d} \phi)^2 \nonumber \\ &&
+ \frac{2M \cos^2 \theta}{\sqrt{\tilde{H}_{1} \tilde{H}_{5}}}[(a_1
c_1 c_5 c_p -a_2 s_1 s_5 s_p) \mathrm{d}t + (a_2 s_1
s_5 c_p - a_1 c_1 c_5 s_p) \mathrm{d}y ] \mathrm{d}\psi \nonumber \\ 
&&+\frac{2M \sin^2 \theta}{\sqrt{\tilde{H}_{1} \tilde{H}_{5}}}[(a_2
c_1 c_5 c_p - a_1 s_1
s_5 s_p) \mathrm{d}t + (a_1
s_1 s_5 c_p - a_2 c_1 c_5 s_p) \mathrm{d}y] \mathrm{d}\phi \nonumber
\\ && + \sqrt{\frac{\tilde{H}_1}{\tilde{H}_5}}\sum_{i=1}^4
\mathrm{d}z_i^2
\end{eqnarray}
where
\begin{eqnarray} 
\tilde{H}_{i}=f+M\sinh^2\delta_i, \quad
f=r^2+a_1^2\sin^2\theta+a_2^2\cos^2\theta,
\end{eqnarray}
and $c_i = \cosh \delta_i$, $s_i = \sinh \delta_i$.  This metric is
more usually written in terms of functions $H_i = \tilde
H_i/f$. Writing it in this way instead makes it clear that there is no
singularity at $f=0$. As the determinant of the metric is
\begin{eqnarray} 
g=-r^2\frac{\tilde{H}_1^3}{\tilde{H}_5}\cos^2\theta\sin^2\theta,
\end{eqnarray}
it is clear that the inverse metric is also regular when
$f=0$. The above metric is in the string frame, and the dilaton is
\begin{equation}
e^{2\Phi} = \frac{\tilde H_1}{\tilde H_5}.
\end{equation}
{}From~\cite{gms1}, the 2-form gauge potential which supports this
configuration is
\begin{eqnarray} \label{rr}
C_2 &=& \frac{M \cos^2 \theta}{\tilde H_1} \left[ (a_2 c_1
  s_5 c_p - a_1 s_1 c_5
  s_p) dt + (a_1 s_1 c_5 c_p - a_2 c_1 s_5 s_p) dy \right] \wedge d\psi
   \\
&& + \frac{M \sin^2 \theta}{\tilde H_1} \left[  (a_1 c_1
  s_5 c_p - a_2 s_1 c_5
  s_p) dt  + (a_2 s_1 c_5 c_p - a_1 c_1 s_5 s_p) dy \right] \wedge d \phi
  \nonumber \\
&& - \frac{M s_1 c_1}{\tilde H_1} dt \wedge dy -
  \frac{M s_5 c_5}{\tilde H_1} (r^2 + a_2^2 + M
  s_1^2) \cos^2 \theta d\psi \wedge d\phi. \nonumber
\end{eqnarray}
We take the $T^4$ in the $z_i$ directions to have volume $V$, and the
$y$ circle to have radius $R$, that is $y \sim y + 2\pi R$. 

Compactifying on $T^4 \times S^1$ yields an asymptotically flat
five-dimensional configuration. The gauge charges are determined by
\begin{equation}
Q_1 = M \sinh \delta_1 \cosh \delta_1,
\end{equation}
\begin{equation}
Q_5 = M \sinh \delta_5 \cosh \delta_5,
\end{equation}
\begin{equation} \label{qp}
Q_p = M \sinh \delta_p \cosh \delta_p,
\end{equation}
where the last is the charge under the Kaluza-Klein gauge field
associated with the reduction along $y$. The five-dimensional Newton's constant is
$G^{(5)} = G^{(10)}/(2\pi R V)$; if we work in units where $4 G^{(5)}/
\pi =1$, the Einstein frame ADM mass and angular momenta are
\begin{equation}
M_{ADM} = \frac{M}{2} (\cosh 2\delta_1 + \cosh 2\delta_5 + \cosh 2 \delta_p ),
\end{equation}
\begin{equation} \label{jpsi}
J_\psi = -  M (a_1 \cosh \delta_1 \cosh \delta_5
  \cosh \delta_p - a_2 \sinh \delta_1 \sinh \delta_5 \sinh \delta_p),  
\end{equation}
\begin{equation} \label{jphi}
J_\phi = -  M (a_2 \cosh \delta_1 \cosh \delta_5
  \cosh \delta_p - a_1 \sinh \delta_1 \sinh \delta_5 \sinh \delta_p)
\end{equation}
(which are invariant under interchange of the $\delta_i$). We see that
the physical range of $M$ is $M \geq 0$. We will assume without loss
of generality $\delta_1 \geq 0$, $\delta_5 \geq 0$, $\delta_p \geq 0$
and $a_1 \geq a_2 \geq 0$.

We also want to rewrite this metric as a fibration over a
four-dimensional base space. It has been shown in~\cite{gmr} that the
general supersymmetric solution in minimal six-dimensional
supergravity could be written as a fibration over a four-dimensional
hyper-K\"ahler base, and writing the supersymmetric two-charge solutions
in this form played an important role in understanding the relation
between these solutions and supertubes in~\cite{lmm} and in an attempt
to generate new asymptotically flat three-charge solutions by spectral
flow~\cite{lunin}. The supersymmetric three-charge solutions were also
written in this form in~\cite{gm}. Of course, in the
non-supersymmetric case, we do not expect the base to have any
particularly special character, but we can still use the Killing
symmetries $\partial_t$ and $\partial_y$ to rewrite the metric
\eqref{3charge} as a fibration of these two directions over a
four-dimensional base space. This gives
\begin{eqnarray} \label{fibred}
\mathrm{d}s^2 &=& \frac{1}{\sqrt{ \tilde{H_1}\tilde{H_5} } }\left\{
-(f-M) \left[\mathrm{d}\tilde{t}- (f-M)^{-1} M
\cosh\delta_1\cosh\delta_5 (a_1 \cos^2\theta \mathrm{d}\psi + a_2 \sin^2\theta 
\mathrm{d}\phi) \right]^2 \right. \nonumber \\ && + f \left.\left[
\mathrm{d}\tilde{y}+f^{-1} M \sinh\delta_1\sinh\delta_5 (a_2
\cos^2\theta \mathrm{d}\psi + a_1 \sin^2\theta \mathrm{d}\phi) \right]^2
\right\}\nonumber\\ && + \sqrt{\tilde{H_1}\tilde{H_5}}\left\{
\frac{r^2\mathrm{d}r^2}{ (r^2+a_1^2)(r^2+a_2^2)-Mr^2 }
+\mathrm{d}\theta^2 \right.\nonumber \\ && + (f(f-M))^{-1}\left[
  \left(f(f-M)+f a_2^2\sin^2\theta - (f-M)a_1^2\sin^2\theta
  \right)\sin^2\theta \mathrm{d}\phi^2 \right. \nonumber \\ && + 2 M
  a_1 a_2 \sin^2 \theta \cos^2 \theta \mathrm{d}\psi \mathrm{d}\phi
  \nonumber \\
  && + \left.\left.\left(f(f-M)+fa_1^2\cos^2\theta -
  (f-M)a_2^2\cos^2\theta\right)\cos^2\theta \mathrm{d}\psi^2
  \right] \phantom{ \frac{1}{1}} \right\},
\end{eqnarray}
where $\tilde t = t \cosh \delta_p - y \sinh \delta_p$, $\tilde y = y
\cosh \delta_p -t \sinh \delta_p $.

We can see that this is still a `natural' form of the metric, even in the
non-supersymmetric case, inasmuch as the base metric in the second \{\} is
independent of the charges. This form of the metric is as a
consequence convenient for studying the `near-horizon' limit, as we
will now see.

In addition to the asymptotically flat metrics written above, we will
be interested in solutions which are asymptotically AdS$_3 \times
S^3$. These asymptotically AdS$_3 \times S^3$ geometries can be
thought of as describing a `core' region in our asymptotically flat
soliton solutions, but they can also be considered as geometries in
their own right. It is relatively easy to identify the appropriate CFT
duals when we consider the asymptotically AdS$_3 \times S^3$
geometries. To prepare the ground for this discussion, we should
consider the `near-horizon' limit in the general family of
metrics.

The near-horizon limit is usually obtained by assuming that $Q_1, Q_5
\gg M, a_1^2, a_2^2$, and focusing on the region $r^2 \ll Q_1, Q_5$. This
limiting procedure is easily described if we consider the metric in
the form \eqref{fibred}: it just amounts to `dropping the 1' in the harmonic
functions $H_1, H_5$, that is, replacing $\tilde H_1 \approx Q_1$,
$\tilde H_5 \approx Q_5$, and also approximating $M \sinh \delta_1
\sinh \delta_5 \approx M \cosh \delta_1 \cosh \delta_5 \approx
\sqrt{Q_1 Q_5}$ in the cross terms in the fibration. This gives us the
asymptotically AdS$_3 \times S^3$ geometry
\begin{eqnarray}
\mathrm{d}s^2 &=& \frac{1}{\sqrt{ Q_1 Q_5 } }\left\{ -(f-M)
[\mathrm{d}\tilde{t}- (f-M)^{-1} \sqrt{Q_1 Q_5} (a_1 \cos^2\theta
\mathrm{d}\psi + a_2 \sin^2\theta \mathrm{d}\phi) ]^2
\right. \nonumber \\ && + f \left.[ \mathrm{d}\tilde{y}+f^{-1}
\sqrt{Q_1 Q_5} (a_2 \cos^2\theta \mathrm{d}\psi + a_1 \sin^2\theta
\mathrm{d}\phi) ]^2 \right\}\nonumber\\ && + \sqrt{Q_1
  Q_5}\left\{ \frac{r^2\mathrm{d}r^2}{ (r^2+a_1^2)(r^2+a_2^2)-Mr^2 }
+\mathrm{d}\theta^2 \right.\nonumber \\ && + (f(f-M))^{-1}\left[
  \left(f(f-M)+f a_2^2\sin^2\theta - (f-M)a_1^2\sin^2\theta
  \right)\sin^2\theta \mathrm{d}\phi^2 \right. \nonumber \\ && + 2 M
  a_1 a_2 \sin^2 \theta \cos^2 \theta \mathrm{d}\psi \mathrm{d}\phi
  \nonumber \\
  && + \left.\left.\left(f(f-M)+fa_1^2\cos^2\theta -
  (f-M)a_2^2\cos^2\theta\right)\cos^2\theta \mathrm{d}\psi^2
  \right]\phantom{ \frac{1}{1}} \right\}. 
\end{eqnarray}
This can be rewritten as
\begin{eqnarray}
\mathrm{d}s^2 &=& - \left(\frac{ \rho^2}{\ell^2} - M_3 +
\frac{J_3^2}{4 \rho^2} \right) d\tau^2 + \left(\frac{ \rho^2}{\ell^2}
- M_3 + \frac{J_3^2}{4 \rho^2} \right)^{-1} d\rho^2 + \rho^2 \left(
d\varphi + \frac{J_3}{2\rho^2} d\tau \right)^2  \nonumber  \\ && + \ell^2 d\theta^2
+
\ell^2 \sin^2 \theta [d\phi + \frac{R}{\ell^2} (a_1 c_p - a_2
s_p) d\varphi + \frac{R}{\ell^3} ( a_2
c_p-a_1 s_p) d \tau]^2 \nonumber \\ 
&&+ \ell^2 \cos^2 \theta [d \psi + \frac{R}{\ell^2}
(a_2 c_p - a_1 s_p) d\varphi + \frac{R}{\ell^3}
(a_1 c_p-a_2 s_p) d \tau]^2,\label{aads}
\end{eqnarray}
where we have defined the new coordinates
\begin{equation}
\varphi = \frac{y}{R}, \quad \tau = \frac{t \ell}{R},
\end{equation}
\begin{equation}
\rho^2 = \frac{R^2}{\ell^2} [r^2 + (M-a_1^2-a_2^2) \sinh^2 \delta_p +
a_1 a_2 \sinh 2\delta_p ]
\end{equation}
and parameters
\begin{equation}
\ell^2 = \sqrt{Q_1 Q_5}, 
\end{equation}
\begin{equation}
M_3 = \frac{R^2}{\ell^4} [ (M-a_1^2-a_2^2) \cosh 2\delta_p + 2 a_1 a_2
  \sinh 2 \delta_p], 
\end{equation}
\begin{equation}
J_3 = \frac{R^2}{\ell^3} [ (M-a_1^2-a_2^2) \sinh 2\delta_p + 2 a_1 a_2
  \cosh 2 \delta_p]. 
\end{equation}
Thus, we see that we recover the familiar observation that the
near-horizon limit of the six-dimensional charged rotating black string is a
twisted fibration of $S^3$ over the BTZ black hole
solution~\cite{clads}.

\section{Finding solitonic solutions}
\label{solitons}

In general, these solutions will have singularities, horizons, and
possibly also closed timelike curves. Let us now consider the
conditions for the spacetime to be free of these features, giving a
smooth solitonic solution. 

Written in the form~\eqref{3charge}, the metric has coordinate
singularities when $\tilde H_1 =0$, $\tilde H_5= 0$ or $g(r) \equiv
(r^2 + a_1^2) (r^2 + a_2^2) - Mr^2 = 0$. In addition, the determinant
of the metric vanishes if $\cos^2 \theta =0$, $\sin^2 \theta=0$, or
$r^2=0$, which will therefore be singular loci for the inverse
metric. The singularities at $\tilde H_1 =0$ or $\tilde H_5= 0$ are
real curvature singularities, so we want to find solutions where
$\tilde H_1 >0$ and $\tilde H_5 > 0$ everywhere. The vanishing of the
determinant at $\theta=0$ and $\theta=\frac{\pi}{2}$ merely signals
the degeneration of the polar coordinates at the north and south poles
of $S^3$; these are known to be just coordinate singularities for
arbitrary values of the parameters, so we will not consider them
further.

The remaining coordinate singularities depend only on $r$. We can
construct a smooth solution if the outermost one is the result of the
degeneration of coordinates at a regular origin in some $\mathbb{R}^2$
factor; that is, of the smooth shrinking of an $S^1$. If this origin
has a large enough value of $r$, we will have $\tilde H_1 >0$ and
$\tilde H_5 > 0$ there, and we will get a smooth solution. The
coordinate singularity at $r^2=0$ cannot play this role, as we can
shift it to an arbitrary position by defining a new radial coordinate
by $\rho^2 = r^2 - r_0^2$. The determinant of the metric in the new
coordinate system will vanish at $\rho^2=0$.

The interesting coordinate singularities are thus those at the roots
of $g(r)$, and the first requirement for a smooth solution is that
this function {\it have} roots. If we write 
\begin{equation}
g(r) = (r^2-r_+^2)(r^2-r_-^2)
\end{equation}
with $r_+^2 > r_-^2$, then
\begin{equation}
r_\pm^2 =\frac{1}{2}(M-a_1^2-a_2^2)\pm \frac{1}{2}\sqrt{
(M-a_1^2-a_2^2)^2-4a_1^2a_2^2 }.
\end{equation}
We see that this function only has real roots for 
\begin{equation}
|M-a_1^2 -a_2^2| > 2a_1 a_2 .
\end{equation}
There are two cases: $M > (a_1+a_2)^2$, or $M < (a_1 - a_2)^2$. Note
that in the former case, $r_+^2 >0$, whereas in the latter, $r_+^2
<0$ (which is perfectly physical, since as noted above, we are free to
define a new radial coordinate by shifting $r^2$ by an arbitrary constant). 

Assuming one of these two cases hold, we can define a new radial
coordinate by $\rho^2 = r^2 - r_+^2$. Since $r^2 dr^2 = \rho^2
d\rho^2$, in this new coordinate system
\begin{equation}
g_{\rho \rho} = \sqrt{ \tilde{H_1}\tilde{H_5} } \frac{d\rho^2}{\rho^2
  + (r_+^2 - r_-^2) },
\end{equation}
and the determinant of the metric is $g =
-\rho^2\frac{\tilde{H}_1^3}{\tilde{H}_5}\cos^2\theta\sin^2\theta$.
Thus, in this coordinate system, the only potential problems are at $\rho^2
= 0$ and $\rho^2 = r_-^2 - r_+^2$, that is, at the two roots of the
function $g(r)$.

To see what happens at $r^2=r_+^2$, consider the geometry of the surfaces
of constant $r$. The determinant of the induced metric is
\begin{equation} \label{fixedr}
g^{(ty\theta\phi\psi)} = - \cos^2 \theta \sin^2 \theta \tilde
H_1^{1/2} \tilde H_5^{1/2} g(r). 
\end{equation}
Thus, at $r^2=r_+^2$, the metric in this subspace degenerates. This can
signal either an event horizon, where the surface $r^2=r_+^2$ is null, or
an origin, where $r^2=r_+^2$ is of higher codimension. We can distinguish
between the two possibilities by considering the determinant of the
metric at fixed $r$ and $t$; that is, in the $(y,\theta,\phi,\psi)$
subspace. This is
\begin{eqnarray} \label{subdet}
g^{(y\theta\phi\psi)} &=& \cos^2\theta \sin^2\theta \left\{ g(r)
\left(r^2+a_1^2\sin^2\theta +a_2^2\cos^2\theta + M(1+s_1^2 +s_5^2+
s_p^2)\right) \right. \nonumber \\ && + r^2 M^2 (c_1^2 c_5^2 c_p^2 -
s_1^2 s_5^2 s_p^2 ) +M^2(M-a_1^2-a_2^2)s_1^2 s_5^2 s_p^2 \nonumber \\
&& \left. + 2M^2a_1a_2 s_1 c_1 s_5 c _5 s_p c_p \right\}.
\end{eqnarray}
This will be positive at $r^2=r_+^2$ if and only if $M > (a_1 +
a_2)^2$. If it is, the constant $t$ cross-section of $r^2=r_+^2$ will
be spacelike, and $r^2=r_+^2$ is an event horizon. Thus,
we can have smooth solitonic solutions without horizons only in the
other case $M < (a_1 - a_2)^2$.

To have a smooth solution, we need a circle direction to be shrinking
to zero at $r^2=r_+^2$. That is, we need some Killing vector with
closed orbits to be approaching zero. Then by a suitable choice of
period we could identify $\rho^2=0$ with the origin in polar
coordinates of the space spanned by $\rho$ and the angular coordinate
corresponding to this Killing vector. The Killing vectors with closed
orbits are linear combinations
\begin{equation} \label{degkv}
\xi = \partial_y - \alpha \partial_\psi - \beta \partial_\phi,
\end{equation} 
so a necessary condition for a circle degeneration is that
\eqref{subdet} vanish at $r^2=r_+^2$, so that some linear combination
of this form has zero norm there. We can satisfy this condition in two
different ways.

\subsection{Two charge solutions: $a_1 a_2=0$}

The first possibility is to set $a_2 = 0$, so $a_1 a_2 = 0$. Then for
$M< a_1^2$, $r_+^2 = 0$, and we set \eqref{subdet} to zero at $r^2=0$
by taking one of the charges to vanish. We will focus on setting
$\delta_p=0$, since these solutions will have a natural interpretation
in CFT terms. Recall that in string theory, we can interchange the
different charges in this solution by U-dualities. 

For this choice of parameters, the metric simplifies to 
\begin{eqnarray}  \label{2chmetric}
\mathrm{d}s^2&=\frac{1}{\sqrt{\tilde{H}_{1}\tilde{H}_{5}}}
\left[-(f-M)(\mathrm{d}t-
(f-M)^{-1}M c_1 c_5 a_1 \cos^2\theta \mathrm{d}\psi)^2\right.
\\&+\left.f(\mathrm{d}y \nonumber
+f^{-1}M s_1 s_5 a_1 \sin^2\theta \mathrm{d}\phi)^2 \right]
\\
&+\sqrt{\tilde{H}_{1}\tilde{H}_{5}}\left(\frac{\mathrm{d}r^2}{r^2+a_{1}^2-M}  
+\mathrm{d}\theta^2+\frac{r^2\sin^2\theta}{f}\mathrm{d}\phi^2
+\frac{(r^2+a_1^2-M)\cos^2\theta}{f-M}\mathrm{d}\psi^2\right). \nonumber
\end{eqnarray}

Since \eqref{subdet} vanishes at $r^2=0$, the orbits of a Killing
vector of the form \eqref{degkv} must degenerate there. It is easy to
use the simplified metric \eqref{2chmetric} to evaluate
\begin{equation}
\alpha = 0, \quad \beta = \frac{a_1}{M s_1 s_5}.
\end{equation}
That is, if we define a new coordinate 
\begin{equation} \label{2chtilde}
\tilde{\phi}=\phi+\frac{a_1}{M s_1 s_5} y,
\end{equation}
the direction which goes to zero is $y$ at fixed $\tilde \phi,
\psi$. To make $y \to y + 2\pi R$ at fixed $\tilde \phi, \psi$ a
closed orbit, we require 
\begin{equation}
\frac{a_1}{M s_1 s_5 } R = m \in \mathbb{Z}.
\end{equation}
Around $r=0$, we then have
\begin{equation}
\mathrm{d}s^2=  \ldots +\sqrt{\tilde{H}_{1}\tilde{H}_{5}} \left(
\frac{\mathrm{d}r^2}{a_{1}^2-M} + \frac{r^2\mathrm{d}y^2}{M^2 s_1^2
s_5^2 } \right)+ \ldots 
\end{equation}
This will be regular if we choose the radius of the $y$ circle to be
\begin{equation} \label{2chperiod}
R =\frac{M s_1 s_5}{\sqrt{a_{1}^2-M}}.
\end{equation}
Thus, the integer quantisation condition fixes 
\begin{equation} \label{2chint}
m=\frac{a_1}{\sqrt{a_{1}^2-M}}.
\end{equation}
With this choice of parameters, the solution is completely smooth, and
$\theta, \tilde \phi, \psi$ are the coordinates on a smooth $S^3$ at
the origin $r=0$. We recover the smooth supersymmetric solutions
of~\cite{bbkr,maldmaoz} for $m=1$.

{}From the CFT point of view, it is natural to regard the charges $Q_1,
Q_5$ and the asymptotic radius of the circle $R$ as fixed
quantities. We can then solve \eqref{2chperiod} and \eqref{2chint} to
find the other parameters, giving us a one integer parameter family of
smooth solutions for fixed $Q_1, Q_5, R$. The integer \eqref{2chint}
determines a dimensionless ratio $a_1^2/M$, while the other condition
\eqref{2chperiod} fixes the overall scale ($a_1$, say) in terms of
$Q_1, Q_5, R$.

\subsection{Three charge solutions}

Solutions with all three charges non-zero can be found by considering
$a_1 a_2 \neq 0$. Setting \eqref{subdet} to zero at $r^2 = r_+^2$
implies that
\begin{equation} \label{mass}
M = a_1^2+a_2^2-a_1 a_2 \frac{(c_1^2 c_5^2 c_p^2 + s_1^2 s_5^2 s_p^2
  )}{s_1 c_1 s_5 c_5 s_p c_p}
\end{equation}
and hence that 
\begin{equation}
r_+^2 =-a_1a_2 \frac{s_1 s_5 s_p}{c_1 c_5 c_p}.
\end{equation}

The Killing vector which degenerates is \eqref{degkv} with\footnote{This
  choice of parameters is most easily derived by requiring $g_{ty} \to
  0$ at $\rho^2 =0$; having derived it, one can then check that it
  also gives $g_{yy} \to 0$ at $\rho^2 =0$ as required.} 
\begin{equation}
\alpha = - \frac{s_p c_p }
{(a_1 c_1 c_5 c_p-a_2 s_1 s_5 s_p)}, \quad \beta = - \frac{s_p c_p }
{(a_2 c_1 c_5 c_p- a_1 s_1 s_5 s_p)}.
\end{equation}
The associated shifts in the $\phi, \psi$ coordinates are hence 
\begin{equation}
\tilde{\psi} = \psi - \frac{s_p c_p }
{(a_1 c_1 c_5 c_p - a_2 s_1 s_5 s_p)} y, \quad \tilde{\phi} = \phi -
\frac{s_p c_p } 
{(a_2 c_1 c_5 c_p - a_1 s_1 s_5 s_p)} y, 
\end{equation}
and $y \to y + 2\pi R$ at fixed $\tilde \phi, \tilde \psi$ will be a
closed orbit if 
\begin{equation}
\frac{s_p c_p }
{(a_1 c_1 c_5 c_p- a_2 s_1 s_5 s_p)} R = n, \quad \frac{s_p c_p } 
{(a_2 c_1 c_5 c_p- a_1 s_1 s_5 s_p)} R = -m 
\end{equation}
for some integers $n,m$.  As in the two-charge case, requiring
regularity of the metric at the origin fixes the radius of the $y$
circle. We do not give details of the calculation, but simply quote
the result,
\begin{equation} \label{period}
R =\frac{ M s_1 c_1 s_5 c_5 (s_1 c_1 s_5 c_5 s_p
  c_p)^{1/2}}{\sqrt{a_1a_2}(c_1^2 c_5^2 c_p^2 - s_1^2 s_5^2 s_p^2 )}.
\end{equation}

If we introduce dimensionless parameters 
\begin{equation}
j = \left( \frac{a_2}{a_1} \right)^{1/2} \leq 1 , \quad s = \left( \frac{ s_1
  s_5 s_p}{ c_1 c_5 c_p} \right)^{1/2} \leq 1, 
\end{equation}
then the integer quantisation conditions determine these via
\begin{equation} \label{integer}
\frac{ j+ j^{-1}}{s + s^{-1}} = m-n, \quad \frac{ j - j^{-1}}{s -
  s^{-1}} = m+n. 
\end{equation}
Note that this constraint is invariant under the permutation of the
three charges. We note that we can rewrite the mass \eqref{mass} as
\begin{equation}
M = a_1 a_2 (s^2 -j^2)(j^{-2} s^{-2} - 1) = a_1 a_2 nm (s^{-2} - s^2)^2,
\end{equation}
so $M \geq 0$ implies $s^2 \geq j^2$ and $nm \geq 0$. Our assumption
that $a_1 > a_2$ implies $n \geq 0$, so $m \geq 0$, and
\eqref{integer} implies $m > n$.

\begin{figure}[htbp]
\centering
\psfrag{aa1}{$\frac{a_1}{\sqrt{M}}$}
\psfrag{aa2}{$\frac{a_2}{\sqrt{M}}$}
\includegraphics[width=0.7\textwidth]{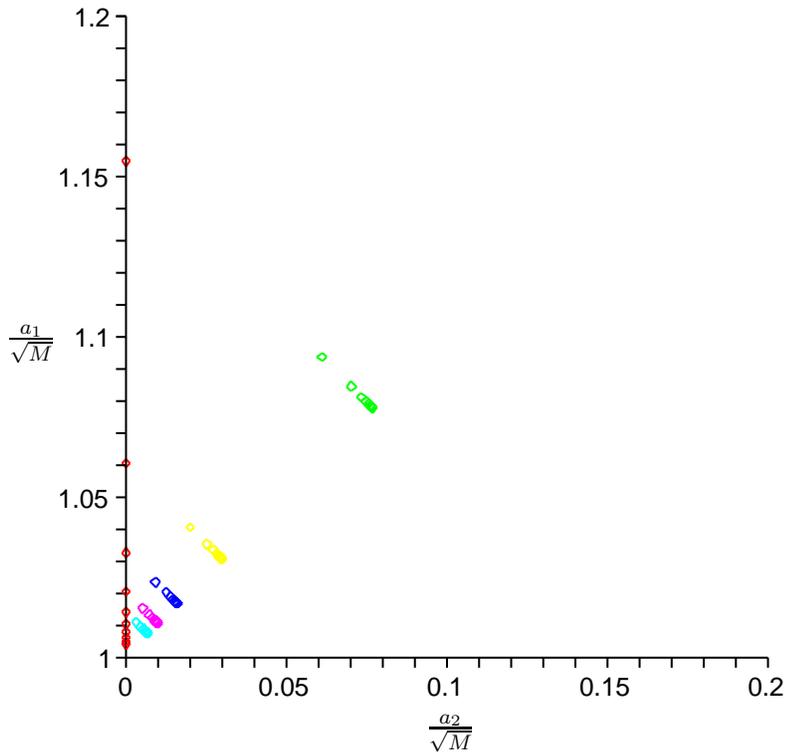}
\caption{The values of the dimensionless quantities $a_2/\sqrt{M},
  a_1/\sqrt{M}$ for which smooth solitons are obtained are indicated
  by points. The highest point on the figure corresponds to $m=2,
  n=0$. Increasing $n$ moves diagonally downwards towards the
  diagonal, and increasing $m-n$ moves down towards $(0,1)$. For each
  point, there is a set of orbifolds labelled by $k$. Solutions with
  event horizons exist in the region $a_1/\sqrt{M} + a_2/\sqrt{M} <
  1$ (off the bottom of the plotted region).}
\label{fig1}
\end{figure}

Thus, in this case, for given $Q_1, Q_5, R$, we have a two integer
parameter family of smooth solutions.  It is a little more difficult
to make direct contact with the supersymmetric solutions
of~\cite{gms1} in this case, since one needs to take a limit $a_1, a_2
\to \infty$, but these would correspond to $m = n+1$, as it turns out
that $s=1$ and $M=0$ if and only if $m = n+1$. We can also think of
the two-charge solutions in the previous subsection as corresponding
to the case $n =0$. To gain some insight into the values of the
parameters for other choices of $m,n$, we have plotted the
dimensionless quantities $a_1/\sqrt{M}, a_2/\sqrt{M}$ for some
representative values in figure \ref{fig1}.

\subsection{Orbifolds \& more general smooth three-charge solution}

So far, we have insisted that the solution be smooth. However, in the
context of string theory, we may also consider solutions with orbifold
singularities, since the corresponding worldsheet conformal field
theory is completely well-defined. In the context of the above smooth
solutions, a particularly interesting class of orbifolds is the
$\mathbb{Z}_k$ quotient by the discrete isometry $(y, \psi, \phi) \sim
(y + 2\pi R/k, \psi, \phi)$. 

In the two-charge case, the quotient acts as $(y, \psi, \tilde \phi)
\sim (y + 2\pi R/k, \psi, \tilde \phi + 2\pi m/k)$ in the coordinates
appropriate near $r=0$. This isometry has a fixed point at $r=0,
\theta=0$, so the resulting orbifold has a $\mathbb{Z}_k$ orbifold
singularity there. In addition, if $k$ and $m$ are not relatively
prime, there will be a $\mathbb{Z}_j$ orbifold singularity at $r=0$
for all $\theta$, where $j = \gcd(k,m)$. The supersymmetric orbifolds
corresponding to $m=1$ have previously been
studied~\cite{bbkr,htube,lm2}.

In the three-charge case, the discrete isometry becomes $(y, \tilde
\psi, \tilde \phi) \sim (y+ 2\pi R/k, \tilde \psi - 2\pi n/k, \tilde
\phi + 2\pi m/k)$, and the $\mathbb{Z}_k$ will be freely acting if $m$
and $n$ are relatively prime to $k$. Thus, we get new smooth
three-charged solutions by orbifolding by a $k$ which is relatively
prime to $m$ and $n$. We could have found such solutions directly if
we had allowed for the possibility that $y \to y + 2\pi R k$ is the
closed circle at $\rho=0$, instead of insisting that it be $y \to y +
2\pi R$. We also have orbifolds similar to the two-charged ones if one
or both of $m$ and $n$ are not relatively prime to $k$. In particular,
the simple supersymmetric orbifolds studied in~\cite{gms2} correspond
to taking $m = kn' +1$, $n = kn'$ for some integer
$n'$.\footnote{In~\cite{gms2}, other examples where $n \neq k n'$ are
obtained by applying STTS duality to these ones. This is possible
because while $(m,n)$ are U-duality
invariant, $k$ is not, so this transformation can map us to new
solutions.} The preserved supersymmetries in the solutions with $m =
n+1$ correspond to Killing spinors which are invariant under
translation in $y$ at fixed $\phi, \psi$, so all the orbifolds of
cases with $m = n+1$ will be supersymmetric. In particular, orbifolds
where $k$ is relatively prime to $n$ and $n+1$ will give new smooth
supersymmetric solutions.

\subsection{Asymptotically AdS solutions}
\label{adssolitons}

In order to understand the dual CFT interpretation of these solutions,
it is interesting to see the effect of the constraints
(\ref{period}, \ref{integer}) on the asymptotically AdS solution
\eqref{aads}. Consider first the two-charge case. If we set $a_2 = 0$,
$\delta_p =0$ and insert (\ref{2chperiod}, \ref{2chint}) in
\eqref{aads}, we will have 
\begin{eqnarray} 
\mathrm{d}s^2 &=& - \left( \frac{\rho^2}{\ell^2} + 1 \right) d\tau^2 +
  \left(  \frac{\rho^2}{\ell^2} + 1 \right)^{-1} d\rho^2 + \rho^2
  d\varphi^2 \nonumber \\
&&+ \ell^2 \left[ d\theta^2 + \sin^2 \theta (d\phi + m
  d\varphi)^2 +  \cos^2 \theta (d \psi + m d\tau/\ell)^2 \right].
\end{eqnarray}
Thus, the asymptotically AdS version of the soliton is just global
AdS$_3 \times S^3$, with a shift of the angular coordinates on the
sphere determined by $m$. 

In the general three-charge case, the interpretation of the dimensionless
parameter $s$ changes in the asymptotically AdS solutions: it is now
$s = \sqrt{\tanh \delta_p}$. The conditions \eqref{integer} are
unaffected, however, and inserting these and the value of the period
\eqref{period} in \eqref{aads}, we will have 
\begin{eqnarray} \label{adsmn}
\mathrm{d}s^2 &=& - \left( \frac{\rho^2}{\ell^2} + 1 \right) d\tau^2 +
  \left( \frac{\rho^2}{\ell^2} + 1 \right)^{-1} d\rho^2 + \rho^2
  d\varphi^2 \\ \nonumber &&+ \ell^2 \left[ d\theta^2 +  \sin^2
  \theta (d\phi + m d\varphi - n d\tau/\ell)^2 +  \cos^2 \theta
  (d \psi - n d\varphi + m d\tau/\ell)^2 \right]. 
\end{eqnarray}
Thus, again, the asymptotically AdS version of the soliton is just global
AdS$_3 \times S^3$, with shifts of the angular coordinates on the
sphere determined by $m,n$.

Thus, in the cases where they have a large `core' region described by
an asymptotically AdS geometry, the smooth solitons studied in the
first two subsections above approach global AdS$_3 \times S^3$ in this
region. As a consequence, the orbifolds studied in the previous
section will have corresponding orbifolds of AdS$_3 \times S^3$; some
of these orbifolds were discussed in~\cite{mmc1,mmc2}.  The resulting
quotient geometry is still asymptotically AdS$_3 \times S^3$, as can
be seen by introducing new coordinates $\varphi' = k \varphi$, $\tau'
= k\tau$, $\rho' = \rho/k$. The metric on the orbifold in these
coordinates is then
\begin{eqnarray} \label{adsorb}
\mathrm{d}s^2 &=& - \left( \frac{\rho'^2}{\ell^2} + \frac{1}{k^2}
  \right) d\tau'^2 + \left( \frac{\rho'^2}{\ell^2} + \frac{1}{k^2} \right)^{-1}
  d\rho'^2 + \rho'^2 d\varphi'^2 \\ \nonumber &&+ \ell^2 \left[ d\theta^2
  +  \sin^2 \theta \left( d\phi + \frac{m}{k} d\varphi' -
  \frac{n}{k \ell} d\tau' \right)^2 \right. \\ \nonumber
&& \left. + \cos^2 \theta \left( d \psi - \frac{n}{k} d\varphi'
  + \frac{m}{k \ell} d\tau' \right)^2 \right].
\end{eqnarray}
The redefined angular coordinate $\varphi'$ will have period $2\pi$ on
the orbifold.

\section{Verifying regularity}
\label{verify}

In the previous section, we claim to have found a family of smooth
solitonic solutions, by imposing three conditions on the parameters of
the general metric. We should now verify that these solutions have no
pathologies. In this section, we will use the radial coordinate
$\rho^2 = r^2 - r_+^2$ (for the two-charge solutions, $\rho^2 = r^2$),
which runs over $\rho \geq 0$.

The first step is to check that $\tilde H_1 >0$, $\tilde H_5 >0$ for
all $\rho \geq 0$, as desired. In these coordinates,
\begin{equation}
f = \rho^2 + (a_1^2-a_2^2) \sin^2 \theta + (a_2^2 -a_1 a_2 s^2).
\end{equation}
In the two-charge case, where $a_2=0$, the last term
vanishes, so $f \geq 0$, and hence $\tilde H_1 >0$, $\tilde H_5 >0$
everywhere. In the more general case, however, the last term is
\begin{equation}
a_2^2 -a_1 a_2 s^2 = -a_1 a_2 (s^2 - j^2) <0,
\end{equation}
so we do not have $f \geq 0$. Examining $\tilde{H}_1$ directly, 
\begin{equation}
\tilde{H}_1 = \rho^2 + (a_1^2-a_2^2) \sin^2 \theta + a_1a_2 ( s^2 -
j^2) (s^{-2} j^{-2} s_1^2 - c_1^2),
\end{equation}
so for $\tilde H_1 >0$ everywhere, we need the last factor to be
positive. We know $s^2 > j^2$, and we can rewrite the last bracket as
\begin{equation}
(s^{-2} j^{-2} s_1^2 - c_1^2) = \frac{c_1^2}{j^2} \left( s^2 \frac{
    c_5^2 c_p^2}{s_5^2 s_p^2} -j^2 \right) >0,
\end{equation}
so we indeed have $\tilde H_1 >0$. We can similarly show $\tilde H_5
>0$. Thus, the metric in the
$(t,\rho,\theta,\tilde{\phi},\tilde{\psi},z^i)$ coordinates is regular
for all $\rho >0$, apart from the coordinate singularities associated
with the poles of the $S^3$ at $\theta =0, \pi/2$, so the local
geometry is smooth. 

Next we check for global pathologies. We can easily see that these
solutions have no event horizons. The determinant of the metric of a
surface of constant $\rho$, \eqref{fixedr}, is negative for $\rho >
0$. That is, there is a timelike direction of constant $\rho$ for all
$\rho > 0$, and hence by continuity there must be a timelike curve
which reaches the asymptotic region from any fixed $\rho$.
We will demonstrate the absence of closed timelike curves by proving a
stronger statement, that the soliton solutions are stably
causal. Using the expression for the inverse metric in appendix
\ref{invmet}, we can evaluate
\begin{equation}
\partial_\mu t \partial_\nu t g^{\mu\nu} = - \frac{1}{\sqrt{ \tilde
    H_1 \tilde H_5} } \left[ f + M (1 + s_1^2 + s_5^2 + s_p^2) +
    \frac{M^2 (c_1^2 c_5^2 c_p^2 - s_1^2 s_5^2 s_p^2)}{\rho^2 + r_+^2
    - r_-^2} \right] < 0,
\end{equation}
so $\partial_\mu t$ is a timelike covector, and $t$ is a global time
function for the solitons. Hence the solitons are stably causal, and
in particular free of closed timelike curves. 

Finally, we should check regularity at $\rho=0$. In the previous
section, we chose $R$ so that the $\rho,y$ coordinates were the polar
coordinates in a smooth $\mathbb{R}^2$. If we define new coordinates
on this $\mathbb{R}^2$ regular at $\rho=0$ by 
\begin{equation}
x^1 = \rho \cos(y/R), \quad x^2 = \rho \sin(y/R), 
\end{equation}
then 
\begin{equation}
dy = \frac{1}{(x_1^2 + x_2^2)} (x^1 dx^2 - x^2 dx^1), 
\end{equation}
and we need the other $g_{\mu y}$ components in the metric to go to
zero at least linearly in $\rho$ for the whole metric to be smooth at
$\rho =0$ once we pass to the Cartesian coordinates $x^1,x^2$. In
fact, we find that the $g_{\mu y}$ go like $\rho^2$ for small $\rho$
in the $(t,\rho,\theta,\tilde{\phi},\tilde{\psi},z^i)$ coordinates.

We also need to verify the regularity of the matter fields. The
dilaton is trivially regular, since $\tilde H_1 >0$, $\tilde H_5 >0$,
but the Ramond-Ramond two-form requires checking. The non-trivial
question is whether the $C_{y \mu}$ go to zero at $\rho^2 =0$. In
fact, in the gauge we used in \eqref{rr}, they don't: we find
\begin{equation}
\begin{aligned}
C_{y \tilde \phi} &= \frac{M s_p c_p s_5 c_5}{a_1 c_1 c_5 c_p - a_2 s_1
  s_5 s_p} + O(\rho^2),\\
C_{y \tilde \psi} &= \frac{M s_p c_p s_5 c_5}{a_2 c_1 c_5 c_p - a_1 s_1
  s_5 s_p} + O(\rho^2), \\
C_{y t} &= \frac{ 1 + s_1^2 + s_p^2}{s_1 c_1} + O(\rho^2). 
\end{aligned}
\end{equation}
We can remove these constant terms by a gauge transformation, so the
Ramond-Ramond fields are regular at $\rho=0$. The physical importance
of the constant terms is that they correspond to electromagnetic
potentials dual to the charges carried by the geometry, and their
presence is presumably related to the first law satisfied by these
soliton solutions, as in~\cite{cvetads}.

In summary, we have shown that the two integer parameter family of
solutions identified in the previous section are all smooth solutions
without CTCs. In the next section, we will explore their relation to
the CFT description of the D1-D5-P system.  

\section{Relation to CFT}
\label{cft}

We have found new smooth solutions by considering the general family
of charged rotating black hole solutions~\eqref{3charge}. These are
labelled by the radius $R$, charges $(Q_1,Q_5)$ and three integers
$(m,n,k)$. They include the previously known supersymmetric solutions
as special cases, and add non-supersymmetric solutions and new
supersymmetric orbifold solutions. We would like to see if we can
relate these solutions to the CFT description, as was done for the
earlier supersymmetric cases in~\cite{bbkr,maldmaoz,gms1}.

If we consider the asymptotically AdS$_3 \times S^3$ solutions
constructed in section \ref{adssolitons}, which describe the `core'
region of the asymptotically flat solitons, we can use the powerful
AdS/CFT correspondence machinery to identify the corresponding states
in the CFT. The dual CFT for the asymptotically AdS$_3 \times S^3
\times T^4$ spaces with radius $\ell = (Q_1 Q_5)^{1/4}$ is a sigma
model with target space a deformation of the orbifold
$(T^4)^N/S_N$~\cite{dcft,swcft,lmcft}, where
\begin{equation}
N= n_1 n_5 = \frac{\ell^4 V}{g^2 l_s^8}, 
\end{equation}
where $V$ is the volume of the $T^4$. This theory has $c = 6 n_1
n_5$. In section \ref{adssolitons}, we showed that the corresponding
asymptotically AdS solutions for a basic family of solitons were
always global AdS$_3 \times S^3$, with a shift on the angular
coordinates on the sphere specified by $n,m$. Following the proposal
outlined in~\cite{bbkr}, we identify the geometries~\eqref{adsmn} with
CFT states with charges
\begin{equation} \label{chs}
\begin{aligned}
h &= \frac{c}{24} (m+n)^2, \quad j &= \frac{c}{12} (m+n)  \\
\bar{h} &= \frac{c}{24} (m-n)^2, \quad \bar{j} &= \frac{c}{12}
(m-n).  
\end{aligned}
\end{equation}
Thus, these states have energy
\begin{equation}
E = h+\bar h = 2(m^2+n^2) \frac{c}{24} = \frac{1}{2} (m^2 +
n^2) n_1 n_5,
\end{equation}
and momentum
\begin{equation} \label{cftmom}
q_p = h - \bar h = 4mn \frac{c}{24} = nm n_1 n_5.
\end{equation}
Since the non-compact geometry is global AdS$_3$, there is a single
spin structure on the spacetime. Because of the shifts in the angular
coordinates, this spin structure can be either periodic or
antiperiodic around $\varphi$ at fixed $\phi, \psi$: it will be
periodic if $m+n$ is odd, and antiperiodic if $m+n$ is even. Thus, the
geometry is identified with a RR state with the above charges if $m+n$
is odd, and with a NSNS state with these same charges if $m+n$ is
even.

These states can be interpreted in terms of spectral flow. Recalling
that spectral flow shifts the CFT charges by~\cite{specflow}
\begin{equation}
h' = h + \alpha j + \alpha^2 \frac{c}{24}, \quad j' = j + \alpha
\frac{c}{12},
\end{equation}
\begin{equation}
\bar h' = \bar h + \beta \bar j + \beta^2 \frac{c}{24}, \quad \bar j'
= \bar j + \beta \frac{c}{12},
\end{equation}
we can see that the required states can be obtained by spectral flow
with $\alpha = m+n$, $\beta = m-n$ acting on the NSNS ground state (for
which $h=j=0$, $\bar h = \bar j = 0$). This spectral flow can be
identified with the coordinate transformation in spacetime which
relates the $(\varphi,\phi,\psi)$ coordinates to the $(\varphi,\tilde
\phi, \tilde \psi)$ coordinates. Thus, we see that the
non-supersymmetric states corresponding to all the geometries labelled
by $m,n$ are constructed by starting with the maximally supersymmetric
NSNS vacuum and applying different amounts of spectral flow. 

In~\cite{bbkr}, the special case $m=1, n=0$ was discussed. In this
case, the spectral flow is by one unit on both the left and the right,
and maps the NS vacuum to a R ground state both on the left and the
right. We can see the supersymmetry of this state from the spacetime
point of view: the covariantly constant Killing spinors in global AdS
have the form 
\begin{equation}
\epsilon^\pm_L = e^{\pm i \frac{\tilde \phi_L}{2}} e^{-i \frac{\varphi}{2}}
\epsilon_0, \quad \epsilon^\pm_R =  e^{\pm i \frac{\tilde \phi_R}{2}} e^{-i
\frac{\varphi}{2}} \epsilon_0,
\end{equation}
so when we shift $\tilde \phi_L = \phi_L + \varphi$, $\tilde \phi_R =
\phi_R +\varphi$, the Killing spinors $\epsilon^+_L$, $\epsilon^+_R$
become independent of $\varphi$, corresponding to the preserved
Killing symmetries in the R ground state. If we consider $m= n+1$, the
spectral flow on the right is by one unit, so $\epsilon^+_R$ is still
independent of $\varphi$. These are the supersymmetric states
considered in~\cite{gms1}, which are R ground states on the right, but
the more general R states obtained by spectral flowing by $2n+1$ units
on the left. Our non-supersymmetric solitons correspond to the more
general non-supersymmetric states obtained by spectral flowing the
NSNS vacuum by $m-n$ units on the right and $m+n$ units on the
left. In~\cite{gms1}, an explicit representation for the R sector
state obtained by spectral flow by $2r+1$ units was given,\footnote{We
use a slightly different notation than~\cite{gms1}.}
\begin{equation}
| 2r+1 \rangle_R = (J^+_{-(2r)})^{n_1 n_5} (J^+_{-(2r-4)})^{n_1 n_5}
  \ldots (J^+_{-2})^{n_1 n_5} | 1 \rangle, 
\end{equation}
where $J^+_{-k}$ is a mode of the $su(2)$ current of the full CFT which
raises $h$ and $j$ by $\Delta h = k$, $\Delta j = 1$, and $|1 \rangle$
is the R ground state with $j = +c/12$ obtained by spectral flow from
the NS ground state. Similarly, one can give an explicit representation
of the NS sector state obtained by spectral flow by $2r$ units,
following~\cite{corr},
\begin{equation}
| 2r \rangle_{NS} = (J^+_{-(2r-1)})^{n_1 n_5} (J^+_{-(2r-3)})^{n_1 n_5}
  \ldots (J^+_{-1})^{n_1 n_5} | 0 \rangle_{NS}. 
\end{equation}
The CFT state corresponding to the geometry~\eqref{adsmn} is then $|
m+n \rangle_R \times | m-n \rangle_R$ or $| m+n \rangle_{NS} \times |
m-n \rangle_{NS}$, depending on the parity of $m+n$.

The situation is more interesting when we consider the orbifolds. The
geometries \eqref{adsorb} should be identified with CFT states with
charges
\begin{equation} \label{orbch}
\begin{aligned}
h &= \frac{c}{24} \left( 1 + \frac{(m+n)^2-1}{k^2} \right), \quad j &=
\frac{c}{12} \frac{m+n}{k},  \\
\bar h &= \frac{c}{24} \left( 1 + \frac{(m-n)^2-1}{k^2} \right), \quad
\bar j &= \frac{c}{12} \frac{m-n}{k}. 
\end{aligned}
\end{equation}
In the supersymmetric case, when $m=n+1$, $\bar h = \frac{c}{24}$,
$\bar j = \frac{c}{12} \frac{1}{k}$, so these geometries still have
the charges of R ground states on the right. This particular R ground
state corresponds to the spectral flow of the NS chiral primary state
with $\bar h = \bar{j} = \frac{c}{24} \frac{k-1}{k}$. However, the
charges of the state in the left-moving sector are, in general, not
those of a R ground state or even the result of spectral flow on a R
ground state. For general $m,n$, neither sector is the spectral flow
of a ground state. Thus, these provide examples of geometries dual to
more general CFT states.

To specify the CFT state completely, we need to say if \eqref{orbch}
are the charges of a RR or a NSNS state. To do so, let us consider the
spin structure on spacetime. When $m$ or $n$ is relatively prime to
$k$, there is a contractible circle in the spacetime, and as a result
the spin structure is fixed. The contractible circle is $(\varphi',
\phi, \psi) \to (\varphi' + 2 \pi k, \phi - 2\pi m, \psi - 2\pi
n)$. The fermions must be antiperiodic around this circle. For the
case where neither $m$ nor $n$ is relatively prime to $k$, we are
not forced to make this choice, but we will assume that we still
choose a spin structure such that the fermions are antiperiodic around
this circle; this would correspond to the spin structure inherited
from the covering space of the orbifold.

In the supersymmetric case $m=n+1$, and more generally for $m+n$ odd,
this implies that the fermions are periodic under $\varphi' \to
\varphi' + 2\pi k$ at fixed $\phi,\psi$. For $k$ odd, this implies the
fermions must be periodic under $\varphi' \to \varphi' + 2\pi$, while
for $k$ even, they may be either periodic or antiperiodic. Thus, for
$m=n+1$, we can always choose the periodic spin structure for the
fermions on spacetime. This spacetime will then be identified with the
supersymmetric RR state with the charges \eqref{orbch}. However, for
$k$ even, we can choose the antiperiodic spin structure for the
fermions on spacetime; this spacetime will then be identified with a
NSNS state with the same charges \eqref{orbch}. In this latter case,
neither the spacetime solution nor the CFT state is supersymmetric.

The situation becomes stranger for $m+n$ even. The antiperiodicity
around the contractible cycle implies that the fermions will be
antiperiodic under $\varphi' \to \varphi' + 2\pi k$ at fixed
$\phi,\psi$. If $k$ is odd, this is compatible with a spin structure
antiperiodic in $\varphi'$, but if $k$ is even, there is no spin
structure on the orbifold which satisfies this condition. The orbifold
cannot be made into a spin manifold. The general conditions for such
orbifolds $M/\Gamma$ to inherit a spin structure from the spin
manifold $M$ were discussed in~\cite{spin}; see also~\cite{quot} for
further discussion relevant to the case at hand. It will be
interesting to see how this obstruction for $k$ even, $m+n$ even is
reflected in the CFT dual.

In the other cases, we can unambiguously identify the CFT state
corresponding to the geometry as the state with charges \eqref{orbch}
in the sector with the same periodicity conditions on the fermions as
in the spacetime (choosing one of the two possible spin structures on
spacetime in the case $k$ even, $m+n$ odd). It would be interesting to
construct an explicit description of these states, as in the
discussion in~\cite{gms1,gms2}.

Thus, there is a clear CFT interpretation of the asymptotically AdS$_3
\times S^3$ geometries. However, the interesting discovery in this
paper is that there are non-supersymmetric asymptotically flat
geometries, and we want to ask to what extent these can also be
identified with individual microstates in the CFT. Clearly the
appropriate CFT states to consider are the ones described above, but
does the identification between state and geometry extend to the
asymptotically flat spacetimes? In particular, does it make sense to
identify the asymptotically flat spacetime with a CFT state in the
general case where it does not have a large approximately AdS$_3
\times S^3$ core region, and there is no supersymmetry?\footnote{The
CFT state for some of the geometries is in the NSNS sector. We do not
regard this as a serious obstruction to an identification at the
classical level: we are considering non-supersymmetric geometries, so
we can allow the fermions to be antiperiodic around the asymptotic
circle in spacetime. At the quantum level, one might worry that these
antiperiodic boundary conditions lead to a constant energy density
inconsistent with the assumed asymptotic flatness.}  We would not in
general expect the match to asymptotically flat geometries to be
perfect, but there is one non-trivial piece of evidence for the
identification of the full asymptotically flat geometries with the CFT
states: the form of the charges still reflects the CFT
structure. Plugging our parameters into (\ref{qp}, \ref{jpsi},
\ref{jphi}) gives
\begin{align}
Q_p &= nm \frac{Q_1 Q_5}{R^2}, \\
J_\phi &= -m \frac{Q_1 Q_5}{R}, \\
J_\psi &= n \frac{Q_1 Q_5}{R}.  
\end{align}
These reproduce the quantisation of the CFT charges in \eqref{chs}. In
the orbifold case, we replace $R$ by $k R$, as the physical period of
the asymptotic circle is $k$ times smaller, and these values now agree
with the charges in \eqref{orbch}. This seems to us like a very
non-trivial consistency check, as it is very difficult to even express
the parameters $M, a_1, a_2$ appearing in the metric \eqref{3charge}
in terms of $Q_1, Q_5$ and $R$ and the integers $m,n$, so there is no
reason why we would have expected to get such a simple result
automatically. So this appears a good reason to believe properties of
the full asymptotically flat geometries are connected to the CFT
states. Note, however, that it does not seem to be possible to cast
the ADM mass in such a simple form. In the next section, we will also
see that the predicted time delay involved in scattering of probes
does not quite match CFT expectations.

\section{Properties of the solitons}
\label{props}

We will briefly discuss some properties of these solutions, and their
relation to the dual CFT. We first discuss the solution of the
massless scalar wave equation in these geometries, following the
discussion in~\cite{htube,lm2,gms2} closely. We then consider the most
significant difference between our non-supersymmetric solitons and the
supersymmetric cases, the absence of an everywhere causal Killing vector. 

\subsection{Wave equation}

It is interesting to study the behaviour of the massless wave equation
on this geometry. This is a first step towards analysing small
perturbations, and also allows us to address questions of scattering
in the geometry which indicate how an exterior observer might probe
the soliton. We consider the massless wave equation on the geometry,
\begin{equation}
\Box \Psi =0.
\end{equation}
It was shown in~\cite{cvetlars} that this equation is separable.
Considering a separation ansatz
\begin{equation} \label{modes}
\Psi = \exp(-i\omega t/R + i \lambda y/R + i m_\psi \psi + i m_\phi \phi)
\chi(\theta) h(r), 
\end{equation}
and using the inverse metric given in appendix \ref{invmet}, we find that the wave equation reduces to 
\begin{equation}
\frac{1}{\sin 2\theta} \frac{d}{d\theta} \left( \sin 2\theta
  \frac{d}{d\theta} \chi \right) + \left[ \frac{(\omega^2 - \lambda^2)}{R^2}
  (a_1^2 \sin^2 \theta + a_2^2 \cos^2 \theta) - \frac{m_\psi^2}{\cos^2
  \theta} - \frac{m_{\phi}^2}{\sin^2 \theta} \right] \chi = -\Lambda \chi,  
\end{equation}
\begin{equation}
\begin{aligned}
&\frac{1}{r} \frac{d}{dr} \left[ \frac{g(r)}{r} \frac{d}{dr} h \right]
  - \Lambda h + \left[ \frac{(\omega^2 - \lambda^2)}{R^2} (r^2 + M s_1^2 + M
    s_5^2) + (\omega c_p + \lambda s_p)^2 \frac{M}{R^2} \right] h
  \\
&-
  \frac{(\lambda - n m_\psi + m m_\phi)^2}{(r^2-r_+^2)} h +
  \frac{(\omega \varrho + \lambda \vartheta - n m_\phi + m
    m_\psi)^2}{(r^2 
    - r_-^2)} h = 0, 
\end{aligned}
\end{equation}
where
\begin{equation}
\varrho = \frac{c_1^2 c_5^2 c_p^2 - s_1^2 s_5^2 s_p^2}{s_1 c_1 s_5
  c_5}, \quad \vartheta = \frac{c_1^2 c_5^2 - s_1^2 s_5^2}{s_1 c_1 s_5
  c_5} s_p c_p.
\end{equation}
We see that the singularity in the wave equation at $r^2= r_+^2$ is
controlled by the frequency around the circle which is shrinking to
zero there. This is a valuable check on the algebra. 
If we introduce a dimensionless variable 
\begin{equation}
x = \frac{r^2 - r_+^2}{r_+^2 - r_-^2}, 
\end{equation}
we can rewrite the radial equation in the form used in~\cite{gms1}, 
\begin{equation}
4 \frac{d}{dx} \left[x (x+1) \frac{d}{dx} h \right]
  + \left( \sigma^{-2} x + 1 - \nu^2 + \frac{\xi^2}{x+1} -
  \frac{\zeta^2}{x} \right) h = 0, 
\end{equation}
where
\begin{align}
\sigma^2 &= \left[ (\omega^2 - \lambda^2) \frac{(r_+^2 - r_-^2)}{R^2}
  \right]^{-1}, \\ 
\nu &= \left[ 1 + \Lambda - \frac{(\omega^2 - \lambda^2)}{R^2} (r_+^2
  + M s_1^2 + M s_5^2) - (\omega c_p + \lambda s_p)^2 \frac{M}{R^2}
  \right]^{1/2}, \\
\xi &= \omega \varrho + \lambda \vartheta -n  m_\phi + m
    m_\psi, \\
\zeta &= \lambda - n m_\psi + m m_\phi. 
\end{align}
We can then use the results of~\cite{gms1}, where the matching of
solutions of this equation in an inner and outer region was carried
out in detail, to determine the reflection coefficient. This
reflection coefficient can be used to determine the time $\Delta t$ it
takes for a quantum scattering from the core region near $x=0$ to
return to the asymptotic region, by expanding $\mathcal{R} = a + b
\sum_n e^{2\pi i n \frac{\omega}{R} \Delta t}$. Their matching procedure
is valid when
\begin{equation}
\sigma^2 \gg 1
\end{equation}
and 
\begin{equation}
\Delta t \gg \frac{R}{(\omega^2 - \lambda^2)^{1/2}}.
\end{equation}
Under these assumptions, their matching procedure gives
\begin{equation}
\Delta t = \pi R_s \varrho,
\end{equation}
where $R_s$ is the radius \eqref{period} for a smooth solution; in the
orbifolds, $R = R_s/k$.  We note that this is in agreement with their
result in the supersymmetric case, as in the limit $\delta_1,
\delta_5, \delta_p \to \infty$,
\begin{equation}
\varrho = \frac{s_1^2 s_5^2 + s_1^2 s_p^2 + s_5^2 s_p^2 + s_1^2 +
  s_5^2 + s_p^2 + 1}{s_1 c_1 s_5 c_5} \approx \frac{Q_1 Q_5 + Q_1 Q_p
  + Q_5 Q_p}{Q_1 Q_5} = \frac{1}{\eta}
\end{equation}
in the notation of~\cite{gms2}. 

In the CFT picture, this travel time is interpreted as the time
required for two CFT modes on the brane to travel around its
worldvolume and meet again. Thus, from the CFT point of view, the
expected value is $\Delta t_{CFT} = \pi R_s$.  As in~\cite{gms2}, there
is a `redshift factor' $\varrho$ between our spacetime result and the
expected answer from the CFT point of view. It was argued
in~\cite{gms2} that such a factor must appear to make the spacetime
result invariant under permutation of the three charges, and it was
proposed that this factor could be understood as a scaling between the
asymptotic time coordinate $t$ in the asymptotically flat space and
the time coordinate appropriate to the CFT. Evidence for this point of
view was found by noting that in the cases where the soliton had a
large AdS$_3 \times S^3$ core region, the global AdS time $\tau$ was
proportional to $\eta t$, so $\Delta \tau = \pi R_s$ in accordance with
CFT expectations. In our non-supersymmetric case, for fixed $m,n$, the
appropriate limit in which we obtain a large AdS region is the limit
$\delta_1, \delta_5 \gg 1$ for fixed $\delta_p$ considered in section
\ref{adssolitons}. We did not see any such scaling between the AdS and
asymptotic coordinates there, but $\varrho \approx 1$ in this limit,
so this is consistent with the interpretation proposed in~\cite{gms2}.
However, we remain uncomfortable with this interpretation. It is hard
to argue directly for such a redshift between the CFT and asymptotic
time coordinates in the general case where the soliton does not have a
large approximately AdS$_3 \times S^3$ core. Indeed, in the dual brane
picture of the geometry, where we have a collection of D1 and D5
branes in a flat background, one would na\"\i vely expect the two to
be the same. A deeper understanding of this issue could shed
interesting light on the limitations of the identification between CFT
states and the asymptotically flat geometries.

\subsection{Ergoregion}
\label{ergo}

Although our soliton solutions are free of event horizons, they
typically have ergoregions. These already appear in the supersymmetric
three-charge soliton solutions studied in~\cite{gms1,gms2}, where the
Killing vector $\partial_t$, which defines time-translation in the
asymptotic rest frame, becomes spacelike at $f=0$ if $Q_p \neq
0$. However, in these supersymmetric cases, there is still a causal
Killing vector (arising from the square of the covariantly constant
Killing spinor), which corresponds asymptotically to the
time-translation with respect to some boosted frame. A striking
difference in the non-supersymmetric solitons is the absence of any
such globally timelike or null Killing vector field.\footnote{For the
asymptotically AdS spacetimes, there is a globally timelike Killing
vector field, given by $\partial_t$ at fixed $\tilde \psi$, $\tilde
\phi$. In $(t, y, \psi, \phi)$ coordinates, this is of the form $V' =
\ell \partial_t - m \partial_\psi + n \partial_\phi$, so it cannot be
extended to a globally timelike Killing vector field in the
asymptotically flat geometry.} The most general Killing vector field
which is causal in the asymptotic region of the asymptotically flat
solutions is
\begin{equation}
V = \partial_t + v^y \partial_y 
\end{equation}
for $|v^y| \leq 1$. However, when $f=0$, the norm of this Killing
vector is
\begin{equation}
|V|^2 = \frac{M}{\sqrt{\tilde H_1 \tilde H_5}} (c_p - v^y s_p)^2 >
 0. 
\end{equation}
 The best we can do is to take $v^y = \tanh \delta_p$, for which this
Killing vector is timelike for $f>M$.  Note that as a consequence, the
two-charge non-supersymmetric solutions also have ergoregions.

In a rotating black hole solution, the existence of an ergoregion
typically implies a classical instability when the black hole is
coupled to massive fields~\cite{bomb,damour}. This instability arises
when we send in a wavepacket which has positive energy less than the
rest mass with respect to the asymptotic Killing time, but negative
energy in the ergoregion. The wavepacket will be partially absorbed by
the black hole, but because the absorbed portion has negative energy,
the reflected portion will have a larger amplitude. This then reflects
off the potential at large distances, and repeats the process. This
process causes the amplitude of the initial wavepacket to grow
indefinitely, until its back-reaction on the geometry becomes
significant.

One might have thought that in the supersymmetric three-charge
solitons, the instability would not appear as a consequence of the
existence of a causal Killing vector, by a mechanism similar to that
discussed in~\cite{hr} for Kerr-AdS black holes. However, this
instability is in fact absent for a different reason, which applies to
both supersymmetric and non-supersymmetric solitons. The instability
in black holes is a result of the existence of both an ergoregion and
an event horizon, so in the solitons, the absence of an event horizon
can prevent such an instability from occurring.  Indeed, from the
discussion of the massless wave equation in the previous section, we
can see that the net flux is always zero, and the amplitude of the
reflected wave is the same as that of the incident wave. That is,
although there is an ergoregion, no superradiant scattering of
classical waves occurs in this geometry, and the mechanism that led to
the black hole bomb does not apply here. There might be an instability
if we considered some interacting theory, as the interactions might
convert part of an incoming wavepacket to negative-energy modes bound
to the soliton, but we will not attempt to explore this issue in more
detail.

Thus, for free fields, there is no stimulated emission at the
classical level. We will now show that there is also no spontaneous
quantum emission.\footnote{We thank Don Marolf for pointing out that
the argument for non-trivial quantum radiation in the original version
of this paper was erroneous, and for explaining the following argument
to us.} There is a natural basis of modes for this geometry; for the
scalar field, \eqref{modes}. To establish which of these
modes are associated with creation and which with annihilation
operators, we need to consider the Klein-Gordon norm 
\begin{equation}
(\Psi, \Psi) = \frac{i}{\hbar} \int_\Sigma d^5 x \sqrt{h} n_\mu
g^{\mu\nu} (\bar \Psi \partial_\nu \Psi - (\partial_\nu \bar \Psi) \Psi),
\end{equation}
where $\Sigma$ is a Cauchy surface, say for simplicity a surface $t =
t_0$, and $n_\mu$ is the normal $n_\mu = \partial_\mu t$. The modes of
positive norm, $(\Psi, \Psi)>0$, correspond to creation operators,
while those of negative norm, $(\Psi, \Psi)<0$, correspond to
annihilation operators. Because of the complicated form of the inverse
metric (see appendix \ref{invmet}), it is difficult to establish
explicitly which are which. However, the main point is that we can
define a vacuum state by requiring that it be annihilated by the
annihilation operators corresponding to all the negative frequency
modes in \eqref{modes}. This will then be the unique vacuum state on
this geometry. Since the modes \eqref{modes} are eigenmodes of both
the asymptotic time-translation $\partial_t$ and of the timelike
Killing vector in the near-core region,  
\begin{equation}
V' = \ell \partial_t - m \partial_\psi + n \partial_\phi,
\end{equation}
these will be the appropriate family of creation and annihilation
operators for observers in both regions. That is, these observers who
follow the orbits of the Killing symmetries will detect no particles
in this state.

Thus, at the level of free fields, the solitons do not suffer from
superradiance at either the classical or quantum level. 

\section{Future directions}
\label{disc}

In this paper, we have found new non-supersymmetric soliton solutions
in the D1-D5 system, and identified corresponding states in the
CFT. These solitons can be viewed as interesting supergravity
backgrounds in their own right. They also provide an interesting
extension of the conjectured identity between CFT microstates and
asymptotically flat spacetimes~\cite{mathurrev,lm2}. 

There are two corresponding classes of issues for further
investigation: further study of the geometry itself, and elucidating
the relation to the dual CFT. In the first category, the classical
stability of these solitons as solutions in IIB supergravity should be
checked. As we discussed in section~\ref{ergo}, although they have
ergoregions, the usual black hole bomb instability will be absent at
least for free fields, as there is no net flux in a scattering off the
geometry. It would be interesting to study stability more generally;
in particular, it would be interesting to know if the geometry suffers
from a Gregory-Laflamme~\cite{gl} type instability if we make the
torus in the $z^i$ directions large.

It would be interesting to try to find asymptotically AdS$_5$
generalizations of these solitons, building on the studies of black
holes in gauged supergravities in~\cite{cvetads}, as in AdS it might
be possible to find non-supersymmetric solitons with a globally
timelike Killing vector. This is known to be possible for some Kerr
black holes in AdS~\cite{hht,hr}.

It would also be interesting to study these solutions as backgrounds
for perturbative string theory. They provide new examples of smooth
asymptotically flat geometries that do not have a global timelike
Killing symmetry, of a rather different character from those presented
in~\cite{bubble}. The existence of supersymmetric special cases may be
a simplifying feature. 

The most important direction of future work to elucidate the relation
of these geometries to the dual CFT is to construct explicit CFT
descriptions of the states dual to the generic orbifold spacetimes and
study their properties from the CFT point of view. The charges for the
dual states found in~\eqref{orbch} show that these states are not
simply the spectral flow of some chiral primary, so they do not
maximize the R-charge for given conformal dimension. They should
therefore be closer to representing the `typical' behaviour of a CFT
state (although they are clearly still very special) and we expect
there will be new tests of the relation between geometry and CFT to be
explored. It will also be interesting to see what happens in the CFT
when we consider the orbifolds with $m+n$ even, $k$ even, where the
spacetime is not a spin manifold.

Another important basic issue from this point of view is to understand
the appearance of stationary geometries dual to non-supersymmetric
states coupled to bulk modes. We would have expected that the CFT
states would decay by the emission of bulk closed string modes. Even
in the simple cases where the near-core geometry is global AdS$_3
\times S^3$, the corresponding CFT state carries comparable numbers of
left and right-moving excitations, which we would expect can interact
to produce bulk gravitons. This physics does not seem to be
represented in our dual geometries. It will be important to study the
decay of these non-supersymmetric states in more detail, and to try to
understand the relation to the soliton.

\section*{Acknowledgements}
We thank Vijay Balasubramanian, Micha Berkooz and Joan Sim\'on for
discussions. We thank Don Marolf for correcting an error in the
previous version, and we thank the participants of the ``Gravitational
aspects of string theory'' workshop for useful comments and
feedback. OM thanks Eileen Madden for hospitality. The work of VJ and
OM is supported by PPARC; the work of SFR and GT is supported by
EPSRC.

\appendix

\section{Inverse metric}
\label{invmet}

To calculate the inverse metric, it is convenient to start from the
fibred form of the metric \eqref{fibred}, construct a corresponding
orthonormal frame, and invert that. For this reason, it is simpler to
give the inverse metric in terms of the boosted coordinates $\tilde t
= t \cosh \delta_p - y \sinh \delta_p$, $\tilde y = y \cosh \delta_p
-t \sinh \delta_p $.

The inverse metric is
\begin{align}
g^{\tilde t\tilde t} &= - \frac{1}{\sqrt{\tilde H_1 \tilde H_5}}
  \left( f + M + M\sinh^2
  \delta_1 + M \sinh^2 \delta_5 + \frac{M^2 \cosh^2 \delta_1 \cosh^2
  \delta_5 r^2}{g(r)} \right), \\ 
g^{\tilde t \tilde y} &= \frac{1}{\sqrt{\tilde H_1 \tilde H_5}} \frac{M^2 \sinh \delta_1
  \sinh \delta_5 \cosh \delta_1 \cosh \delta_5 a_1 a_2}{g(r)}, \\ 
g^{\tilde t\phi} &= -\frac{1}{\sqrt{\tilde H_1 \tilde H_5}} \frac{M \cosh \delta_1
  \cosh \delta_5 a_2 (r^2+a_1^2)}{g(r)}, \\ 
g^{\tilde t\psi} &= -\frac{1}{\sqrt{\tilde H_1 \tilde H_5}} \frac{M \cosh \delta_1
  \cosh \delta_5 a_1 (r^2+a_2^2)}{g(r)}, \\ 
g^{\tilde y \tilde y} &= \frac{1}{\sqrt{\tilde H_1 \tilde H_5}} \left(
  f + M\sinh^2 \delta_1
  + M \sinh^2 \delta_5 + \frac{M^2 \sinh^2 \delta_1 \sinh^2 \delta_5 (r^2 +
  a_1^2 + a_2^2 -M)}{g(r)} \right) , \\
g^{\tilde y\phi} &= -\frac{1}{\sqrt{\tilde H_1 \tilde H_5}}\frac{M \sinh \delta_1
  \sinh \delta_5 a_1 (r^2+a_1^2-M)}{g(r)}, \\ 
g^{\tilde y\psi} &= -\frac{1}{\sqrt{\tilde H_1 \tilde H_5}} \frac{M \sinh \delta_1
  \sinh \delta_5 a_2 (r^2+a_2^2-M)}{g(r)}, \\ 
g^{rr} &= \frac{1}{\sqrt{\tilde H_1 \tilde H_5}} \frac{g(r)}{r^2}, \\
g^{\theta \theta} &= \frac{1}{\sqrt{\tilde H_1 \tilde H_5}}, \\
g^{\phi\phi} &= \frac{1}{\sqrt{\tilde H_1 \tilde H_5}} \left(
  \frac{1}{\sin^2 \theta} 
  + \frac{(r^2+a_1^2)(a_1^2-a_2^2)-M a_1^2}{g(r)} \right) , \\
g^{\phi\psi} &= -\frac{1}{\sqrt{\tilde H_1 \tilde H_5}} \frac{Ma_1
  a_2}{g(r)}, \\ 
g^{\psi\psi} &= \frac{1}{\sqrt{\tilde H_1 \tilde H_5}}\left(
  \frac{1}{\cos^2 \theta} 
  + \frac{(r^2+a_2^2)(a_2^2-a_1^2)-M a_2^2}{g(r)} \right). 
\end{align}

\bibliographystyle{utphys}  
 
\bibliography{solitons}

\end{document}